\documentclass[submission,copyright,creativecommons]{eptcs}
\providecommand{\event}{GandALF 2017} 
\usepackage{underscore}           
\usepackage{amsmath,amssymb,bussproofs,epsfig,graphicx,verbatim}
\usepackage[roscoe]{csp-cm}
\usepackage[all]{xy}
\usepackage{galois}
\usepackage{color}
\usepackage{hhline}
\usepackage{multirow}
\usepackage{tikz}
\usetikzlibrary{arrows,automata}

\newcommand{\circled}[1]{\raisebox{.5pt}{\textcircled{\raisebox{-.9pt} {#1}}}}

\newcommand{\sqsub}{\,\raisebox{-.5ex}
  {$\stackrel{\textstyle\sqsubset}{\scriptstyle{\sim}}$}\,}

\newcommand{\aasg}{\,\raisebox{0.065ex}{:}{=}\,}
\newcommand{\sbr}[1]{\lbrack \! \lbrack #1 \rbrack \! \rbrack}
\newcommand{\san}[1]{\langle #1 \rangle}
\newcommand{\ibr}[1]{[ #1 \rangle}

\newcommand{\sq}{\textit{q}}
\newcommand{\sr}{\textit{run}}
\newcommand{\sok}{\textit{ok}}
\newcommand{\srd}{\textit{read}}
\newcommand{\sd}{\textit{done}}
\newcommand{\sw}{\textit{write}}
\newcommand{\sa}{\textit{abort}}

\newcommand{\impif}[4][]{\mbox{$\mathsf{if}$}^{#1} ~\ensuremath{{#2}~ \mbox{$\mathsf{then}$} ~{#3}~ \mbox{$\mathsf{else}$} ~{#4}}}

\newcommand{\Nat}{{\mathbb N}}
\newcommand{\true}{\texttt{true}}

\newtheorem{theorem}{Theorem}

\newtheorem{example}[theorem]{Example}
\newtheorem{proposition}[theorem]{Proposition}

\title{Probabilistic Analysis Based On Symbolic Game Semantics and Model Counting }%
\author{Aleksandar S. Dimovski
\institute{Computer Science Department, IT University of Copenhagen, Copenhagen, Denmark }
\email{adim@itu.dk}
}

\def\titlerunning{Probabilistic Analysis Based On Symbolic Game Semantics and Model Counting}
\def\authorrunning{Aleksandar S. Dimovski}
\begin{document}
\maketitle

\begin{abstract}

Probabilistic program analysis aims to quantify the probability that a given program satisfies a required property.
It has many potential applications, from program understanding and debugging to
computing program reliability, compiler optimizations and quantitative information flow analysis for security.
In these situations, it is usually more relevant to quantify the probability of satisfying/violating
a given property than to just assess the possibility of such events to occur.

In this work, we introduce an approach for probabilistic analysis of open programs (i.e.\ programs
with undefined identifiers) based on game semantics and model counting.
We use a symbolic representation of algorithmic game semantics to collect the symbolic constraints
on the input data (context) that lead to the occurrence of the target events (e.g.\ satisfaction/violation of a given property).
The constraints are then analyzed to quantify how likely is an input to satisfy them.
We use model counting techniques to count the number of solutions (from a bounded integer domain)
that satisfy given constraints.
These counts are then used to assign probabilities to program executions and to assess the probability for the target event to occur at the
desired level of confidence.
Finally, we present the results of applying our approach to several interesting examples and
illustrate the benefits they may offer.

\end{abstract}

\section{Introduction}\label{sec:introduction}

In order to understand program behaviour better, apart from finding out whether a behaviour (execution) can
successfully terminate or not, we often need to know how \emph{likely} a behaviour is to occur.
In particular, we want to distinguish between what is possible behaviour (even with extremely low probability)
and what is likely behaviour (possible with higher probability).
In this work, we show how to calculate the probability of 
behaviours and estimate
the reliability of programs by using a combination of (symbolic) game semantics and model counting.

\emph{Game semantics} \cite{DBLP:journals/entcs/AbramskyM96,DBLP:journals/iandc/HylandO00}
 is a technique for building models of programs that are \emph{fully abstract}, i.e.\
sound and complete with respect to observational equivalence.
The notion of observational equivalence relies on comparing the outcomes of placing programs
in all possible syntactic contexts (environments).
Its algorithmic subarea \cite{GM,DL,DBLP:conf/sas/DimovskiGL05,DBLP:journals/fmsd/KieferMOWW13} aims to apply game semantics
models to software verification by providing concrete automata-based representations for them.
The key characteristics of game semantics models are the following.
They provide precise and compact summaries of observable (input and output) program behaviour,
without showing the explicit reference to a state (state manipulations are hidden).
There is a model for any open program with free (undefined) identifiers
such as calls to library functions.
Finally, the models are generated
inductively (compositionally) on the structure of programs, which is often essential for the \emph{modular} analysis
of larger programs.
Symbolic representation
of game semantics models \cite{D14} extends the (standard) regular-language representation \cite{GM}
by using symbolic data values instead of concrete ones for the inputs.
This allows us to obtain compact models
of programs by using finite-state symbolic automata.
Each \emph{complete symbolic play} (accepting word) in the model  corresponds to a program execution (path), and it is guarded
by a conjunction of constraints on the symbols, known as \emph{play condition}, which indicate under what conditions
this play (word, execution) is feasible.
If the play condition is satisfied by some concrete values for symbols, then they represent input values
that will allow the execution to follow the specific path through the code.
For the generation of symbolic game models where each play is associated with a play condition we
use the \textsc{Symbolic GameChecker} \footnote{\url{https://aleksdimovski.github.io/symbolicgc.html}.} tool \cite{D14}. 
\emph{Model counting} is the problem of determining the number of solutions of a given constraint (formula).
The \textsc{LattE} \footnote{\url{http://www.math.ucdavis.edu/~latte}. UC Davis, Mathematics. } tool \cite{LattE} implements state-of-the-art algorithms for computing volumes, both real and integral,
of convex polytopes as well as integrating functions over those polytopes.
In particular, we use model counting techniques and the \textsc{LattE} tool to estimate
algorithmically the exact number of points of a bounded (possibly very large) discrete domain
that satisfy given linear constraints.

In this paper, we describe a method based on symbolic game models and model counting for performing
a specific type of quantitative analysis -- the calculation of play probabilities and the program reliability.
Calculating the probability of a symbolic play (path) involves counting the number of solutions to the play condition (by using model counting),
and dividing it by the total space of values of the inputs (context).
We assume that the input values are \emph{uniformly distributed} within their finite discrete domain.
We label each (complete) symbolic play with either \emph{success} or \emph{failure} depending on whether a designated
$\mathsf{abort}$ command is executed or not.
Since the set of play conditions produced by the
symbolic game model is a complete partition of the given finite input domain, we can compute the reliability
of the program as the probability of satisfying any of the successful play conditions.
To account for cycles (infinite behaviours) in the model, we use bounded analysis.
For a ``$\mathsf{while}$'' command, a bound is set for the exploration depth (i.e.\ the number of re-visited states).
For an undefined (first-order) function, we restrict the number of times the function
can call its arguments when placed in the given bounded context, thus obtaining a finite input domain.

The main contributions of this work are:
(1) A demonstration of how to add path probabilities using (symbolic) algorithmic game semantics and model counting;
(2) An application of our approach to calculate the reliability of open programs;
(3) A prototype implementation as part of \textsc{Symbolic GameChecker}. 


\section{Programming Language} \label{sec:language}

The use of meta-languages is very common in the semantics community. The semantic model
is defined for a meta-language, and a real programming language (C, ML, etc.) can be studied
by translating it into this meta-language and using the induced model.
Here we consider Idealized Algol (IA), a well studied meta-language 
introduced by Reynolds \cite{Reynolds}.
IA enables functional (typed call-by-name $\lambda$-calculus) and imperative programming.
For the purpose of obtaining an automata-based representation of game semantics,
we shall consider its second-order recursion-free fragment
(IA$_2$ for short). Its types are: 
\[
D ::= \mathsf{int} \, \mid \, \mathsf{bool} \qquad B ::= \mathsf{exp} D \, \mid \, \mathsf{com} \, \mid \, \mathsf{var} D \qquad T ::= B \mid B \rightarrow T
\]
where $D$, $B$, and $T$ stand for data types, base types, and first-order function types, respectively.
The \emph{syntax} of the language is:
\[
\begin{array}{l}
M ::= \! x \! \mid \! v \! \mid \! \mathsf{skip} \! \mid \! \mathsf{diverge} \! \mid \! M\, \mathsf{op} \, M \mid M ; \!\! M \mid \!
 \mathsf{if}\,M\,\mathsf{then}\,M \,\mathsf{else}\, M \! \mid \! \mathsf{while}\,M\,\mathsf{do}\,M   \\
 \qquad \ \mid  M := M \mid \, !M \mid \mathsf{new}_D \: x\!:=\!v \: \mathsf{in} \: M
  \mid \mathsf{mkvar}_D MM \mid \! \lambda x . M \mid MM
\end{array}
\]
where $x$ ranges over a countable set of identifiers, and $v$ ranges over constants of type $D$, which includes integers ($n$) and booleans ($tt, ff$).
The standard arithmetic-logic operations $\mathsf{op}$ are employed, as well as
the usual imperative constructs: sequential composition ($;$), conditional ($\mathsf{if}$), iteration ($\mathsf{while}$),
assignment ($:=$), de-referencing operator ($!$) which is used for reading the value stored in a variable,
a ``do-nothing'' command ($\mathsf{skip}$), and a divergence command ($\mathsf{diverge}$).
Block-allocated local variables are introduced by a $\mathsf{new}$ construct, which initializes a variable and makes it local
to a given block. They are also called ``good'' (storage) variables since what is read from a variable is the last value written into it.
The construct $\mathsf{mkvar}$ is used for creating so-called ``bad'' variables, which do not behave like genuine storage variables \cite{DBLP:conf/lics/HarmerM99}.
There are also standard functional constructs for function definition and application.
\emph{Well-typed terms} 
are given by typing judgements of the form $\Gamma
\vdash M : T$, where $\Gamma=x_1:T_1, \ldots, x_k:T_k$ is a type \emph{context} consisting of
a finite number of typed free identifiers.
Typing rules are given in \cite{DBLP:journals/entcs/AbramskyM96,Reynolds}.

The \emph{operational semantics} is defined by a big-step reduction
relation:
\[
\Gamma \vdash M,\mathrm{s} \Longrightarrow V,\mathrm{s}'
\]
where $\Gamma \vdash M:T$ is a term in which all free identifiers from $\Gamma$ are variables,
i.e.\ $\Gamma=x_1:\mathsf{var} D_1, \ldots, x_k:\mathsf{var} D_k$,
and $\mathrm{s}$, $\mathrm{s}'$ represent the \emph{state} before and
after reduction. The state is a function
assigning data values to the variables in $\Gamma$.
Canonical forms (values) are defined by
$V ::= \, x \, \mid \, v  \, \mid \, \lambda x. M
\, \mid \, \mathsf{skip} \, \mid \, \mathsf{mkvar}_{D}MN$.
Reduction rules are standard
(see \cite{DBLP:journals/entcs/AbramskyM96,Reynolds} for details).
Given a closed term $\vdash M : \mathsf{com}$, which has no free identifiers, we say that
$M$ \emph{terminates} if
 $\vdash M, \emptyset \Longrightarrow
\mathsf{skip},\emptyset$.
We define a \emph{program context} $C[-]:\mathsf{com}$ to be a term with zero or more holes $[-]$ in it,
such that if $\Gamma\vdash M:T$ is a term of the same type as the hole then $C[M]$ is a well-typed
closed term of type $\mathsf{com}$, i.e.\ $\vdash C[M] : \mathsf{com}$.
We say that
a term $\Gamma\vdash M:T$ is an \emph{approximate}
of a term $\Gamma\vdash N:T$, written $\Gamma\vdash M \sqsub N$, if and
only if for all contexts $C[-]:\mathsf{com}$, such that
$\vdash C[M]:\mathsf{com}$ and $\vdash C[N]:\mathsf{com}$,
 if $C[M]$ terminates  then $C[N]$ terminates.
If two terms approximate each other they are considered
\emph{observationally-equivalent}, denoted by
$\Gamma\vdash M \cong N$.
In general, observational equivalence is very difficult to reason about due to
the universal quantification over all syntactic contexts $C[-]$ in which the terms
can be placed. 

\section{Symbolic Game Models} \label{sec:symbolic}

We now give a brief overview of symbolic representation of the algorithmic game semantics
for IA$_2$ \cite{D14}.
Let $Sym$ be a countable set of symbolic names, ranged over by
$X$, $Y$, $Z$.
For any finite $W \subseteq Sym$, the function $new(W)$ returns
a minimal symbolic name which does not occur in $W$, and sets $W:=W \cup \{new(W)\}$.
A minimal symbolic name not in $W$ is the one which occurs earliest
in a fixed enumeration of all possible symbolic names.
Let $Exp$ be a set of expressions, ranged over by $e$, generated
by data values ($v \in D$), symbols ($X \in Sym$), and arithmetic-logic
operations ($\mathsf{op}$). We use $a$ to range over arithmetic expressions ($AExp$) and $b$ over
boolean expressions ($BExp$).

Let $\mathcal{A}$ be an alphabet of letters.
We define a \emph{symbolic alphabet} $\mathcal{A}^{sym}$ induced by $\mathcal{A}$ as follows:
\[
\mathcal{A}^{sym} = \mathcal{A} \cup \{ ?X, e \mid X \in Sym, e \in Exp \}
\]
The letters of the form $?X$ are called \emph{input symbols}.
They represent a mechanism for dynamically generating new symbolic names. 
More specifically, $?X$ creates a stream of fresh symbolic names, binding $X$ to the next symbol
from its stream, $new(W)$, whenever $?X$ is evaluated (met).
We use $\alpha$ to range over $\mathcal{A}^{sym}$.
Next we define a \emph{guarded alphabet} $\mathcal{A}^{gu}$ induced by
$\mathcal{A}$ as the set of pairs of boolean conditions and symbolic letters:
\[
\mathcal{A}^{gu} = \{ \ibr{b, \alpha} \mid b \in BExp, \alpha \in \mathcal{A}^{sym} \}
\]
A guarded letter $\ibr{b, \alpha}$ is $\alpha$ only if $b$ evaluates to true otherwise it is the constant $\emptyset$
(the language of $\emptyset$ is $\emptyset$), i.e.
$if \, (b=tt) \, then \, \alpha \, else \, \emptyset$.
We use $\beta$ to range over $\mathcal{A}^{gu}$.
We will often write only $\alpha$ for the guarded letter $\ibr{tt, \alpha}$.
A word $\ibr{b_1, \alpha_1} \cdot \ibr{b_2, \alpha_2} \ldots \ibr{b_n, \alpha_n}$
over $\mathcal{A}^{gu}$ can be represented as a pair $\ibr{b, w}$,
where $b=b_1 \land b_2 \land \ldots \land b_n$ is a boolean condition
and $w=\alpha_1 \cdot \alpha_2 \ldots \alpha_n$ is a word of symbolic letters.

We now describe how IA$_2$ terms can be translated into
symbolic regular languages and symbolic automata. 
Each type $T$ is interpreted by an alphabet of moves $\mathcal A_{\sbr{T}}$ defined as follows:
\[
\begin{array}{l}
 \mathcal{A}_{\sbr{\mathsf{exp}D}} = \{ \sq \} \cup \mathcal{A}_{\sbr{D}}, \
   \mathcal{A}_{\sbr{\mathsf{com}}} = \{ \sr, \sd \}, \
 \mathcal{A}_{\sbr{\mathsf{var}D}} = \{ \sw(a), \srd, \sok, a  \mid  a
\in \mathcal{A}_{\sbr{D}} \} \\
 \mathcal{A}_{\sbr{B_1^{\san{1}} \to \ldots \to B_k^{\san{k}} \to B}}^{gu} = \displaystyle{\sum_{1 \leq i \leq k}} \mathcal{A}_{\sbr{B_i}}^{gu \, \san{i}} + \mathcal{A}_{\sbr{B}}^{gu}
\end{array}
\]
where $\mathcal{A}_{\sbr{\mathsf{int}}} = \mathbb{Z}$, $\mathcal{A}_{\sbr{\mathsf{bool}}} = \{ tt, ff \}$, and
$+$ denotes a disjoint union of alphabets.
Function types are tagged by a superscript $\san{i}$ to keep
record from which type, i.e.\ which component of the disjoint union, each move comes from.
The letters in the alphabet $\mathcal A_{\sbr{T}}$ represent the
\emph{moves}, i.e.\ observable actions that a term of type $T$ can perform.
Each move is either a \emph{question} (a demand for
information) or an \emph{answer} (a supply of information).
For expressions in $\mathcal{A}_{\sbr{\mathsf{exp}D}}$, there is a \emph{question} move
$\sq$ to ask for the value of the expression, and values from $\mathcal{A}_{\sbr{D}}$ to \emph{answer} the question.
For commands, there is a \emph{question} move
$\sr$ to initiate a command, and an \emph{answer} move $\sd$ to signal
successful termination of a command.
For variables, there are \emph{question} moves for writing to the variable, $\sw(a)$, which are
acknowledged by the \emph{answer} move $\sok$; and there is a \emph{question} move $\srd$ for reading from the variable,
which is \emph{answered} by a value from $\mathcal{A}_{\sbr{D}}$.

For any term, we define
a (symbolic) regular-language which represents its game semantics,
i.e.\ its set of complete symbolic plays.
A \emph{play} is a sequence of moves played by two players in turns:
$\textbf{P}$ (Player) which represents the term being modeled, and $\textbf{O}$ (Opponent) which
represents its context. Every (complete) symbolic play
represents the observable effects
of a completed execution (path) of the given term.
It is given as a guarded word $\ibr{b,w}$, where
$b$ is also called the \emph{play condition}.
Assumptions about a symbolic play to be feasible are
recorded in its play condition.
For infeasible plays, the play condition is unsatisfiable,
thus no assignment of concrete values to symbolic names exists that makes
the play condition true.
The regular expression  for $\Gamma \vdash M : T$, denoted as $\sbr{\Gamma \vdash M : T}$,
is defined over the guarded alphabet:
\[\mathcal A_{\sbr{\Gamma \vdash T}}^{gu} = \big( \sum_{x : T' \in
\Gamma} \mathcal{A}_{\sbr{T'}}^{gu \, \san{x}} \big) +
\mathcal{A}_{\sbr{T}}^{gu}
\]
where moves corresponding to types of free identifiers
are tagged with their names to indicate the origin of moves. Hence, 
$\sbr{\Gamma \vdash M : T}$ contains only
observable moves associated with types of free identifiers from $\Gamma$ (suitably tagged) as well
as moves of the top-level type $T$.

The representation of constants is standard:
\[
 \sbr{\Gamma \! \vdash \! v \! : \! \mathsf{exp}D} \! = \! \sq \cdot v \qquad
 \sbr{\Gamma \! \vdash \! \mathsf{skip} \! : \! \mathsf{com}} \! = \! \sr \cdot \sd \qquad  \sbr{\Gamma \! \vdash \! \mathsf{diverge} \! : \! \mathsf{com}} \! = \! \emptyset
\]
For example, an  integer or boolean constant $v$ is modeled by a play where the initial question $\sq$
(``what is the value of this expression?'') is answered by the value of that constant $v$.

Free identifiers are represented by the
so-called copy-cat regular expressions,
which contain all possible behaviours of terms of that type, thus providing the most general context for an open term.
Thus,
\begin{align} \label{eq:free-fun}
& \sbr{\Gamma, x:\mathsf{exp}D_1^{\san{x,1}} \!\to\! \ldots \mathsf{exp}D_k^{\san{x,k}} \!\to \mathsf{exp}D^{\san{x}} \! \vdash \!  x : \mathsf{exp}D_1^{\san{1}} \!\to\! \ldots \mathsf{exp}D_k^{\san{k}} \!\to \mathsf{exp}D} \qquad \qquad \notag \\
& \qquad \qquad \qquad \qquad \qquad \qquad \qquad \qquad = \sq \cdot \sq^{\san{x}} \cdot \big( \sum_{1 \leq i \leq k} \sq^{\san{x,i}} \cdot \sq^{\san{i}} \cdot ?Z_i^{\san{i}} \cdot Z_i^{\san{x,i}} \big)^* \cdot ?X^{\san{x}} \cdot X
\end{align}
When a call-by-name non-local function $x$ with $k$ arguments is called, it may evaluate any of
its arguments, zero or more times, in an arbitrary order (hence, the Kleene closure *)
and then it returns any allowable answer $X$ from its result type.
Recall that the input symbol $?Z$ creates a stream of fresh symbolic names for each instantiation of $?Z$.
Thus, whenever $?Z$ is met in a play,
the mechanism for fresh symbol generation is used to dynamically instantiate it with a new fresh symbolic name from its stream, which binds all occurrences of $Z$
that follow in the play until a new $?Z$ is met which overrides the previous symbolic name with the next symbolic name taken from its stream.
For example, consider the term $f : \mathsf{expint}^{\san{f,1}} \to \mathsf{expint}^{\san{f,2}} \to \mathsf{expint}^{\san{f}} \vdash f : \mathsf{expint}^{\san{1}} \to \mathsf{expint}^{\san{2}} \to \mathsf{expint}$,
where $f$ is an undefined function with two arguments. Its symbolic model is:
\begin{equation} \label{eq:free-f}
\sq \cdot \sq^{\san{f}} \cdot \big( \sq^{\san{f,1}} \cdot \sq^{\san{1}} \cdot ?Z_1^{\san{1}} \cdot Z_1^{\san{f,1}} + \sq^{\san{f,2}} \cdot \sq^{\san{2}} \cdot ?Z_2^{\san{2}} \cdot Z_2^{\san{f,2}} \big)^* \cdot ?X^{\san{f}} \cdot X
\end{equation}
The play corresponding to function ``$f$'' which evaluates its first argument two times,
after instantiating its input symbols $?Z_1$ and $?X$
is given as: $\sq \cdot \sq^{\san{f}} \cdot \sq^{\san{f,1}} \cdot \sq^{\san{1}} \cdot Z_{1,1}^{\san{1}} \cdot Z_{1,1}^{\san{f,1}} \cdot
\sq^{\san{f,1}} \cdot \sq^{\san{1}} \cdot Z_{1,2}^{\san{1}} \cdot Z_{1,2}^{\san{f,1}} \cdot X^{\san{f}} \cdot X$,
where $Z_{1,1}$ and $Z_{1,2}$ are two different symbolic names used to denote values
of the first argument when it is evaluated the first and the second time, respectively.
Therefore, we are using the streaming symbol $?Z_1$ to create different symbolic names so that we can produce
distinct values (independent from one another) if $?Z_1$ is evaluated multiple times during the execution.
Note that letters tagged with $\san{f}$ represent the actions of calling and
returning from the function $f$, while letters tagged with $\san{f,1}$ (resp. $\san{f,2}$) are the actions caused
by evaluating the first (resp. second) argument of $f$.

\begin{table}[t]
\fbox{
\begin{minipage}{100ex}
\begin{tabbing}
 $\sbr{\mathsf{op} : \mathsf{exp}D_1^{\san{1}} \times \mathsf{exp}D_2^{\san{2}} \to \mathsf{exp}D } =
 \sq \cdot \sq^{\san{1}} \cdot ?Z^{\san{1}} \cdot \sq^{\san{2}} \cdot ?Z'^{\san{2}} \cdot (Z \, \mathsf{op} \, Z')$ \\

 $\sbr{\mathsf{;} : \mathsf{com}^{\san{1}} \times \mathsf{com}^{\san{2}} \to \mathsf{com} } =
  \sr \cdot \sr^{\san{1}} \cdot \sd^{\san{1}} \cdot \sr^{\san{2}} \cdot \sd^{\san{2}} \cdot \sd$ \\

 $\sbr{\mathsf{if} : \mathsf{expbool}^{\san{1}} \times \mathsf{com}^{\san{2}} \times \mathsf{com}^{\san{3}} \to \mathsf{com} } =
  \ibr{tt, \sr }  \cdot \ibr{tt, \sq^{\san{1}} }  \cdot \ibr{tt, ?Z^{\san{1}} } \cdot$ \\
 $\qquad \qquad \qquad \qquad \qquad \qquad \qquad \big( \ibr{Z, \sr^{\san{2}} } \cdot \ibr{tt, \sd^{\san{2}} } + \ibr{\neg Z, \sr^{\san{3}} } \cdot \ibr{tt, \sd^{\san{3}} } \big) \cdot \ibr{tt, \sd }$ \\

 $\sbr{\mathsf{while} : \mathsf{expbool}^{\san{1}} \times \mathsf{com}^{\san{2}} \to \mathsf{com} } =
   \ibr{tt, \sr }  \cdot \ibr{tt, \sq^{\san{1}} }  \cdot \ibr{tt, ?Z^{\san{1}} } \cdot$ \\
  $\qquad \qquad \qquad \qquad \qquad \qquad \qquad \big( \ibr{Z, \sr^{\san{2}} } \cdot \ibr{tt, \sd^{\san{2}} } \cdot \ibr{tt, \sq^{\san{1}} }  \cdot \ibr{tt, ?Z^{\san{1}} } \big)^{*} \cdot \ibr{\neg Z, \sd} $ \\

 $\sbr{\mathsf{:=} : \mathsf{var}D^{\san{1}} \times \mathsf{exp}D^{\san{2}} \to \mathsf{com} } =
   \sr \cdot \sq^{\san{2}} \cdot ?Z^{\san{2}} \cdot \sw(Z)^{\san{1}} \cdot \sok^{\san{1}} \cdot \sd$ \\

 $\sbr{\mathsf{!} : \mathsf{var}D^{\san{1}} \to \mathsf{exp}D } =
  \sq \cdot \srd^{\san{1}} \cdot ?Z^{\san{1}} \cdot Z$ \\

 $\mathsf{cell}_{v}^{\san{x}} = (\ibr{?X\!\!=\!\!v,\srd^{\san{x}}} \cdot X^{\san{x}})^* \cdot \big( \sw(?X)^{\san{x}}
 \cdot \sok^{\san{x}} \cdot (\srd^{\san{x}} \cdot X^{\san{x}})^*
 \big)^*$
\end{tabbing}
\end{minipage}
}
\vspace{-0.5mm}
\caption{Symbolic representations of some language constructs} \label{csp.rl2}
\end{table}

The representations of some language constructs ``$\mathsf{c}$'' are given in Table~\ref{csp.rl2}.
Observe that letter conditions different than $tt$ occur only in plays
corresponding to ``$\mathsf{if}$'' and ``$\mathsf{while}$'' constructs.
In the case of ``$\mathsf{if}$'' construct, when the value of the first argument given by the symbol $Z$ is true
then its second argument is run, otherwise if $\neg Z$ is true
then its third argument is run. A composite term $\mathsf{c}(M_1, \ldots, M_k)$ built out
of a language construct ``$\mathsf{c}$'' and subterms $M_1, \ldots, M_k$ is interpreted
by composing the regular expressions for $M_1, \ldots, M_k$ and the regular expression for ``$\mathsf{c}$''.
For example, we have:
\[
\sbr{\Gamma \vdash \impif{B}{M}{M'}:\mathsf{com}} = \sbr{\Gamma \vdash B:\mathsf{expbool}^{\san{1}} } \comp \sbr{\Gamma \vdash M:\mathsf{com}^{\san{2}} } \comp \sbr{\Gamma \vdash M':\mathsf{com}^{\san{3}} } \comp \sbr{\mathsf{if}}
\]
where $\sbr{\mathsf{if}:\mathsf{expbool}^{\san{1}} \times \mathsf{com}^{\san{2}} \times \mathsf{com}^{\san{3}} \to \mathsf{com}}$ is defined in Table~\ref{csp.rl2}.
Composition of regular expressions ($\comp$) is defined as ``parallel composition followed by hiding'' in CSP style \cite{DBLP:journals/entcs/AbramskyM96}. The parallel composition is matching (synchronizing) of the moves in the shared types, whereas hiding is
deleting of all moves from the shared types.
Conditions of the shared (interacting) moves (guarded letters) in the composition are conjoined, along with the condition
that their symbolic letters are equal \cite{D14}.
The $\mathsf{cell}_{v}^{\san{x}}$ regular expression in Table~\ref{csp.rl2} is used to impose the good variable
behaviour on a local variable $x$ introduced using $\mathsf{new}_D \: x\!:=\!v \: \mathsf{in} \: M$. Note that $v$ is
the initial value of $x$, and $X$ is a symbol used to track the current value of $x$.
The $\mathsf{cell}_{v}^{\san{x}}$ behaves as a storage cell and plays
the most recently written value in $x$ in response to $\srd$, or if no value
has been written yet then answers $\srd$ with the initial value $v$.
The model $\sbr{\mathsf{new}_D \: x\!:=\!v \: \mathsf{in} \: M}$ is obtained by constraining the model of $M$,
$\sbr{\mathsf{var}_D \: x \vdash M}$, only
to those plays where $x$ exhibits good variable behaviour described by $\mathsf{cell}_{v}^{\san{x}}$, and then by deleting (hiding)
all moves associated with $x$ since $x$ is local and so not visible outside of the term \cite{D14}.

The following formal results are proved before \cite{D14}.
We define an \emph{effective alphabet} of a regular expression to be the set of all letters
that appear in the language denoted by that regular expression.
The effective alphabet of a regular expression representing any term $\Gamma \vdash M:T$
contains only a \emph{finite subset} of letters from $\mathcal A_{\sbr{\Gamma \vdash T}}^{gu}$, which
includes all constants, symbols, and expressions used for interpreting free identifiers, constructs,
and local variables in $M$.
\begin{proposition}
For any IA$_2$ term, the set $\sbr{\Gamma \vdash M : T}$
is a (symbolic) regular-language without infinite summations defined over its effective finite alphabet.
Moreover, a finite-state symbolic automata $\mathcal{A} \sbr{\Gamma \vdash M : T}$
which recognizes it is effectively constructible.
\end{proposition}


Suppose that there is a special free identifier $\mathsf{abort}$ of type $\mathsf{com}$.
We say that a term $\Gamma \vdash M$ is \emph{safe} 
\footnote{$M[N/x]$ denotes the capture-free substitution of $N$ for $x$ in $M$.} iff
$\Gamma \vdash M[\mathsf{skip}/\mathsf{abort}] \sqsub M[\mathsf{diverge}/\mathsf{abort}]$;
otherwise we say that a term is \emph{unsafe}.
We say that one play is \emph{safe} if it does not contain moves from $\mathcal A_{\sbr{\mathsf{com}}}^{\san{\sa}}$;
otherwise we say that the play is \emph{unsafe}.

\begin{proposition} \label{safe}
A term $\Gamma \vdash M:T$ is safe
iff all plays in $\sbr{\Gamma \!\vdash\! M\!:\!T}$ are safe.
\end{proposition}
For example, $\sbr{\mathsf{abort:com^{\san{abort}}} \vdash \mathsf{skip \, ; abort:com }}
= \sr \, \cdot \, \sr^{\san{abort}} \, \cdot \, \sd^{\san{abort}} \, \cdot \, \sd $, so this term is unsafe
since its model contains an unsafe play.

\begin{example}\label{exp:M}
Consider the term $M$:
\[
\begin{array}{l}
n : \mathsf{expint}^{n}, \mathsf{abort}:\mathsf{com}^{abort} \vdash \ \mathsf{new_{int}} \, x:=0 \ \mathsf{in} \
  \mathsf{while} \, (!x<n) \ \mathsf{do} \ x:=!x+1; \\
\qquad \qquad \qquad \qquad \qquad \qquad \qquad \qquad \qquad \ \ \mathsf{if} \, (!x>1) \ \mathsf{then} \ \mathsf{abort} : \mathsf{com}
\end{array}
\]
\noindent
The model for this term is given in Fig.~\ref{warm2} \footnote{For simplicity, in examples we omit to write angle brackets $\san{,}$ in superscript tags of moves.}.
The dashed edges indicate moves of the environment ($\textbf{O}$) and solid edges
moves of the term ($\textbf{P}$). They serve only as a visual aid to the reader. Accepting states are designated by an interior circle.
Observe that the term communicates with its environment using non-local
identifiers $n$ and $\mathsf{abort}$. So in the model will only be represented
actions associated with $n$ and $\mathsf{abort}$ as well as with the top-level type $\mathsf{com}$.
The input symbol $?X$ is used to keep track of the current value of the local variable $x$
(note that $X$ occurs only in conditional part of plays).
Each time the term ($\textbf{P}$) asks for a value of $n$ with the move $\sq^{n}$, the
environment ($\textbf{O}$) provides a new fresh symbol $?N$ for it.
Note that we consider all possible environments (contexts) in which a term can be placed. Therefore,
the undefined expression $n$ may obtain different value at each call in the above term \cite{GM}.
At this point, the term ($\textbf{P}$) has three possible options
depending on the current values of symbols $N$ and $X$: it can terminate successfully with $\sd$;
it can execute $\mathsf{abort}$ and terminate; or it can run the assignment $x \aasg x+1$ and ask
for a new value of $n$.

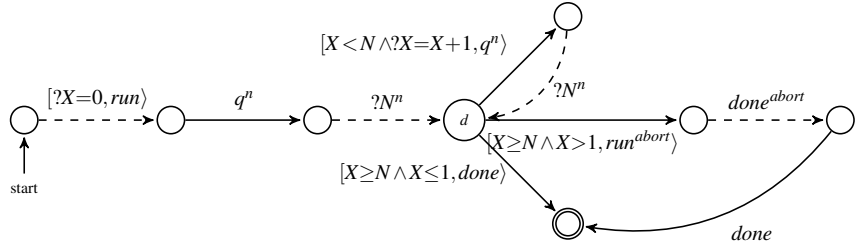
\begin{figure}[t]
\begin{center}
\begin{tikzpicture}[->,>=stealth',shorten >=1pt,auto,node distance=1.95cm,
                    semithick]
  \tikzstyle{every state}=[minimum size=.25pt,initial text={{\tiny start}}]

  \node[initial below,state] (A)                    {};
  \node[state]         (B) [right of=A] {};
  \node[state]         (C) [right of=B] {};
  \node[state]         (D) [right of=C]  {${\scriptscriptstyle d}$};
  \node[state,accepting]         (E) [below right of=D]  {};
  \node[state]         (F) [above right of=D]  {};
  \node[state]         (H) [right of=D, node distance=30.5mm]  {};
  \node[state]         (I) [right of=H]  {};

  \path[font=\scriptsize,dashed] (A) edge node {$\ibr{?X\!\!=\!\!0,\sr}$} (B)
        (C) edge              node {$?N^n$} (D)
                (H) edge              node[above] {$\sd^{\sa}$} (I)
        (F) edge [bend left=40] node[right] {$?N^n$} (D);
  \path[font=\scriptsize]
        (B) edge              node {$\sq^n$} (C)
        (D) edge              node {$\ibr{X\!<\!N \!\land\! ?X\!\!=\!\!X\!+\!1, \sq^n}$} (F)
        (D) edge              node[below] {$\ibr{X\!\!\geq\!\!N \!\land\! X\!\!>\!\!1, \sr^{\sa}}$} (H)
        (D) edge              node[below,left] {$\ibr{X\!\!\geq\!\!N \!\land\! X\!\!\leq\!\!1, \sd}$} (E)
        (I) edge [bend left]  node {$\sd$} (E);
\end{tikzpicture}
\end{center}
\vspace{-2mm}
 \caption{The symbolic game model for $M$. }
\label{warm2}
\end{figure}

\end{example}

\section{Calculating Success and Failure Probabilities}

In this section, we define the success and failure probability of terms, and show how they can
be automatically calculated using symbolic game models and model counting.
We also show how to cope with cases that introduce infinite behaviours.

\subsection{Definition}

We define the \emph{success probability} as the probability that a term terminates successfully
without hitting any failure, such as running the $\mathsf{abort}$ command.
On the other hand, the \emph{failure probability} is the probability that a term hits a failure
during its execution.
The resulting symbolic game model is a set of symbolic plays (words), each with a play condition.
Some of these plays are unsafe (i.e.\ lead to a failure, abortion); whereas some of them are safe (i.e.\ lead to a successful termination without abortion).
The plays are therefore classified in two sets: $P^s$ which contains safe plays, and $P^f$ which contains unsafe plays.

Our discussion focusses on the case of computing probabilities for
terms that have \emph{finite input domains} for all their plays (executions).
This is achieved by constraining all identifiers from $\Gamma$ to be of types in which
only finite sets of basic data values $D$ are used.
For example, we may consider only the basic types over $\mathsf{bool}$ and $\mathsf{int}_k = [0,k)=\{0,\ldots,k-1\}$
for any $k>0$. We also need to bound the input domain when
undefined (first-order) functions are used. This case is handled separately in Section~\ref{bound:ana}.
Finally, we restrict our attention on play conditions expressed as
\emph{linear integer arithmetic} (LIA) constraints
over symbols whose values are \emph{uniformly distributed} over their finite input domain.

Given a symbolic play $p \in \sbr{\Gamma \vdash M:T}$, let $ID_p$ be the total space of possible values in
 its finite input domain and let $pc_p$ be its play condition (constraint).
We now show how to calculate the probability of $p$ occurring, denoted $Pr(p)$.
We use
the \textsc{LattE} tool to compute the number of elements of $ID_p$
that satisfy $pc_p$, denoted $\#(pc_p)$.
The size of $ID_p$, denoted $\#(ID_p)$, is the product of domain's sizes of all symbols
instantiated in $p$, which correspond to all calls of free identifiers of types in which data values $D$ are used.
Thus, we have:
$\#(ID_p) = \prod_{Z \in p} |\!dom(Z)\!|$ and $Pr(p)=\#(pc_p) / \#(ID_p)$, where $|\!dom(Z)\!|=k$ if $Z$ is a symbol
that represents a value from the finite domain $\mathsf{int}_k$.
Note that the size of the input domain (context) $ID_p$ for each play $p$ can be different, and depends on how many
symbols have been instantiated in $p$ that correspond to the data type $D$.
The play conditions associated with plays from $P^s$ and $P^f$ define disjoint input sets and cover the whole finite input domain,
thus defining a complete partition of the finite input domain.
Finally, we define
the \emph{success probability} (resp., \emph{failure probability}) as the probability of evaluating the term $\Gamma \vdash M:T$
within a context (input) that enables all safe (resp., unsafe) plays:
\begin{equation} \label{for:success}
\begin{array}{l}
Pr^s(\Gamma \!\vdash\! M\!:\!T) = \sum_{p \in P^s} \frac{ \#(pc_p)}{\#(ID_{p})}, \quad
 Pr^f(\Gamma \vdash M:T) = \sum_{p \in P^f} \frac{ \#(pc_p)}{\#(ID_{p})}
\end{array}
\end{equation}
Note that $Pr^s(\Gamma \vdash M:T) + Pr^f(\Gamma \vdash M:T) =1$.

\begin{example} \label{exp:prob}
Consider the term $M'$:
\[
n : \mathsf{expint}_{10}^{n}, abort:\mathsf{com}^{abort}  \vdash
 \texttt{if} \ (n \geq 5) \ \texttt{then} \ \mathsf{skip} \ \texttt{else} \ \mathsf{abort} \mathsf{: com}
 \]
Its symbolic game model is:
\[
\sr \cdot \sq^{n} \cdot ?N^{n} \cdot \big(\ibr{N\!\geq\!5, \sd} + \ibr{N\!<\!5, \sr^{\sa}} \cdot \sd^{\sa} \cdot \sd \big)
\]
Suppose that $n \in [0,10)$ and that the possible values for $n$ are independently and uniformly distributed
across this range.
Thus, after instantiation of the input symbol $?N$, there are one safe play ($\sr \cdot \sq^{n} \cdot N^{n} \cdot \ibr{N\!\geq\!5, \sd}$)
and one unsafe play ($\sr \cdot \sq^{n} \cdot N^{n} \cdot \ibr{N\!<\!5, \sr^{\sa}} \cdot \sd^{\sa} \cdot \sd$).
The safe (resp., unsafe) play condition is: $N \geq 5$ (resp., $N<5$). Thus, we obtain $Pr^s(M')=5/10 \, (50\%)$ and $Pr^f(M')=5/10 \, (50\%)$.
\end{example}

We use model counting and the \textsc{LattE} tool \cite{LattE} to determine the number of solutions
of a given constraint. \textsc{LattE} accepts LIA constraints expressed as
a system of linear inequalities each of which defines a hyperplane encoded as the matrix inequality:
$Ax \leq B$, where $A$ is an $m \times n$ matrix of coefficients and $B$ is an $n \times 1$ column vector
of constants. Most LIA constraints can easily be converted into the form: $a_1x_1 + \ldots + a_nx_n \leq b$.
For example, $\geq$ and $>$ can be flipped by multiplying both sides by $-1$, and strict inequalities $<$
can be converted by decrementing the constant $b$.
In \textsc{LattE} equalities $=$ can be expressed directly.
If we have disequalities  $\neq$, they can be handled by counting a set of constraints that encode
all possible solutions. For example, the constraint $\alpha \land (x_1 \neq x_2)$ is handled by finding
the sum of solutions for $\alpha \land (x_1 \leq x_2-1)$ and $\alpha \land (x_1 \geq x_2+1)$.
For a system $Ax \leq B$, where $A$ is an $m \times n$ matrix and $B$ is an $n \times 1$ column vector,
the input \textsc{LattE} file is:
\[
\begin{matrix}
m \ & \ n\!+\!1 \\
B \ & \ -A
\end{matrix}
\]

For example, the constraint ``$N<5$'' from Example~\ref{exp:prob}
results in the following (hyperplane) H-representation for \textsc{LattE}:
\[
\begin{array}{lc}
3 \ & 2 \ \\
9 \ & -1 \\
0 \ & 1 \\
4 \ & -1
\end{array}
\]
where the first line indicates the matrix size: the number of inequalities by the number of variables plus one.
The next two inequalities encode the max and min values for the symbol $N$ based on its data type.
The last inequality expresses the constraint: $N \leq 4$ (i.e.\ $N<5$).
\textsc{LattE} reports that there are exactly 5 points that satisfy the above inequalities ($N \leq 9 \land N \geq 0 \land N<5$).

\subsection{Bounded Analysis} \label{bound:ana}

The presence of ``$\mathsf{while}$'' command and free identifiers of function
type (i.e.\ undefined functions) introduce infinite behaviors, a cycle, in our model.
Hence, convenient analysis strategies are required for handling them in order to compute the
success and failure probabilities.
In the case of the ``$\mathsf{while}$'' command, the source of infinite behaviour is the \emph{term}
being modeled, but the context is still finite. On the other hand, in the case of undefined
(first-order) functions, the source of infinite behaviour is the \emph{context} in which
that function can be placed (e.g. the function may call its arguments infinitely many times), so
the context is unbounded in this case. This is the reason why we have two different strategies
to cope with ``$\mathsf{while}$'' and undefined functions.

\paragraph{The $\mathsf{while}$ command.}
The solution is based on bounded exploration:
a (user-defined) \emph{bound} $d \in \Nat$ is set for the search depth (i.e.\ the number of times a state can be re-visited).
When the bound is reached the search backtracks.
Intuitively, the bound $d \in \Nat$ represents the number of iterations of the $\mathsf{while}$-loop and so we have
the following bounded definition for $\mathsf{while}$ (instead of the one in Table~\ref{csp.rl2}):
\[
\begin{array}{l}
\sbr{\mathsf{while} : \mathsf{expbool}^{\san{1}} \times \mathsf{com}^{\san{2}} \to \mathsf{com} } = \ibr{tt, \sr }  \cdot \ibr{tt, \sq^{\san{1}} }  \cdot \ibr{tt, ?Z^{\san{1}} } \cdot \qquad \qquad \\
   \qquad \qquad \qquad \qquad \qquad \qquad
    \sum_{k=0}^{d} \big( \ibr{Z, \sr^{\san{2}} } \cdot \ibr{tt, \sd^{\san{2}} } \cdot \ibr{tt, \sq^{\san{1}} }  \cdot \ibr{tt, ?Z^{\san{1}} } \big)^{k} \cdot \ibr{\neg Z, \sd}
\end{array}
\]

In this setting the search is no longer
complete, and besides safe and unsafe plays, a new set of plays is collected for
traces interrupted before completing the search.
We call this set of plays \emph{grey} and label it as $P^g$.
We can define $Pr^g(\Gamma \vdash M:T)$ analogously to the other sets as shown in Eqn.~(\ref{for:success}).
The three sets of play conditions associated with plays in $P^s$, $P^f$, and $P^g$ are disjoint and constitute a complete partition
of the entire finite input domain. Hence, $Pr^s(\Gamma \vdash M:T) + Pr^f(\Gamma \vdash M:T) + Pr^g(\Gamma \vdash M:T) =1$.
The intuitive meaning of $Pr^g(\Gamma \vdash M:T)$ is to quantify the plays of $\sbr{\Gamma \vdash M:T}$ for which neither
safety nor unsafety have been revealed at the current exploration depth.
This information is a measure of the \emph{confidence} we can put on our success (resp., failure)
estimation obtained within the given exploration bound: $Confidence = 1 - Pr^g(P)$.
$Confidence=1$ means that the search is complete, i.e.\ for each input we can state if
it leads to a safe or an unsafe execution. Increasing the exploration depth, the confidence grows
revealing more accurate safe (resp., unsafe) predictions.

\begin{example}
Let us reconsider the term $M$ from Example~\ref{exp:M}.
Suppose that $n$ is of type $\mathsf{expint}_{10}$. 
We will now calculate the values of $Pr^s$, $Pr^f$, $Pr^g$, and $Confidence$, for different exploration depths $d$.
Let $d=0$. This means the state $\circled{{\small d}}$ from its symbolic model given in Fig.~\ref{warm2} can be visited only once
(i.e.\ $\circled{{\small d}}$ cannot be re-visited).
Let $N$ be the symbol name instantiated for $?N$.
In this case, there is one unsafe play: $\ibr{X\!\!=\!\!0,\sr} \cdot \sq^n \cdot N^n \cdot \ibr{X\!\!\geq\!\!N \!\land\! X\!\!>\!\!1, \sr^{\sa}}
\cdot \sd^{\sa} \cdot \sd$, and one safe play: $\ibr{X\!\!=\!\!0,\sr} \cdot \sq^n \cdot N^n \cdot \ibr{X\!\geq\!N \!\land\! X\!\!\leq\!1, \sd}$.
The condition of the unsafe play is unsatisfiable (note $X\!\!=\!\!0 \!\land\! X\!\!>\!\!1$) and so $Pr^f(M)=0$; whereas the condition of the safe play is satisfiable with
only one solution for $N=0$ and so $Pr^s(M)=1/10 \, (10\%)$.
For $N \in [1,10)$, the state $\circled{{\small d}}$ needs to be re-explored so $Pr^g(M)=9/10$ and
 $Confidence=1/10 \, (10\%)$.

Let $d=1$. This means the state $\circled{{\small d}}$ in Fig.~\ref{warm2} can be re-visited once.
Let $N_1$ and $N_2$ be the symbol names instantiated when $?N$ is evaluated the first and the second time, respectively.
In this case, there are two unsatisfiable unsafe plays and two safe plays.
The first safe play is from the previous iteration corresponding to $N_1=0$ with probability $1/10$.
The second safe play is: $\sr \cdot \sq^n \cdot N_1^n \cdot \sq^n \cdot N_2^n \cdot \sd$, with the condition:
$X_1\!=\!0 \land X_1 \!<\! N_1 \land X_2\!=\!X_1\!+\!1 \land X_2 \!\geq\! N_2 \land X_2 \!\leq\! 1$,
which has 18 solutions: for $N_1 \in [1,10)$ and $N_2 \in [0,2)$.
Thus, $Pr^s(M)=1/10 + (9/10) \cdot (2/10) = 28/100 \, (28\%)$; $Pr^f(M)=0$; $Pr^g(M)=72/100$;
and $Confidence=28/100 \, (28\%)$.

Let $d=2$ and let $N_1$, $N_2$, $N_3$ be the symbol names instantiated the first, the second, and the third time when $?N$ is met, respectively.
We obtain unsafe plays when $N_1 \in [1,10)$, $N_2 \in [2,10)$, and $N_3 \in [0,3)$, and so we have
$Pr^f(M) \!=\! (9/10) \cdot (8/10) \cdot (3/10) \!=\! 21.6\%$,
$Pr^s(M)\!=\!28\%$, $Pr^g(M)\!=\!50.4\%$, and $Confidence\!=\!49.6\%$.
For $d=3$, we have
$Pr^f(M)\!=\!41.76\%$, $Pr^s(M)\!=\!28\%$, $Pr^g(M)\!=\!30.24\%$, and $Confidence\!=\!69.76\%$.
\end{example}

\paragraph{Undefined functions.}
Recall the definition of undefined functions in Eqn.~(\ref{eq:free-fun}).
The `generic behaviour' of a call-by-name function is, when called by its context, to
perform some sequence of calls to its arguments, and then to return a result.
Since the number of times the function's arguments are called can be arbitrary (even infinite,
see the Kleene closure in Eqn.~(\ref{eq:free-fun})), the
corresponding input domain is not finite. One solution is to place numeric bounds on the number of times
an undefined function can call its arguments.
For any integer $d > 0$, we define $\Gamma \vdash_{d} M:T$ as a term which can be placed into
contexts where any of its first-order free identifiers from $\Gamma$ can call its arguments
at most $d-1$ times.

For example, the interpretation of
$f : \mathsf{expint}^{\san{f,1}} \to \mathsf{expint}^{\san{f,2}} \to \mathsf{expint}^{\san{f}} \vdash f : \mathsf{expint}^{\san{1}} \to \mathsf{expint}^{\san{2}} \to \mathsf{expint}$
now becomes:
\begin{equation} \label{eq:free-f2}
\sq \cdot \sq^{\san{f}} \cdot \textstyle \sum_{k=0}^{d-1} \big( \sq^{\san{f,1}} \cdot \sq^{\san{1}} \cdot ?Z_1^{\san{1}} \cdot Z_1^{\san{f,1}} + \sq^{\san{f,2}} \cdot \sq^{\san{2}} \cdot ?Z_2^{\san{2}} \cdot Z_2^{\san{f,2}} \big)^{k} \cdot ?X^{\san{f}} \cdot X
\end{equation}
Thus, we now use the bound $d$ instead of the Kleene closure *, which is used in the general case given in Eqn.~(\ref{eq:free-f}).
Let us calculate the sizes of input domains corresponding to individual plays from the above model in Eqn.~(\ref{eq:free-f2}).
Assume that we work with the finite integer domain $\mathsf{int}_{10}=[0,10)$.
The play $p_0=\sq \cdot \sq^{\san{f}} \cdot X^{\san{f}} \cdot X$ corresponds to a function
``$f$'' which does not evaluate its arguments at all (a non-strict function), and so there are 10 different instantiations
of $p_0$  since $X \in [0,10)$. Note that if the play condition is $\true$, which means that all
instantiations of $p_0$ are feasible then $\#(\true_{p_0})=10$.
If ``$f$'' evaluates its arguments once, then we have two plays: $p_{1,1}=\sq \cdot \sq^{\san{f}} \cdot \sq^{\san{f,1}} \cdot \sq^{\san{1}} \cdot Z_1^{\san{1}} \cdot Z_1^{\san{f,1}} \cdot
X^{\san{f}} \cdot X$ (``$f$'' evaluates its first argument) with $10^2$ different instantiations corresponding to $Z_1,X \in [0,10)$, and
$p_{1,2}=\sq \cdot \sq^{\san{f}} \cdot \sq^{\san{f,2}} \cdot \sq^{\san{2}} \cdot Z_2^{\san{2}} \cdot Z_2^{\san{f,2}} \cdot
X^{\san{f}} \cdot X$ (``$f$'' evaluates its second argument) with $10^2$ different instantiations corresponding to $Z_2,X \in [0,10)$.
For a function ``$f$'' that calls its arguments $d-1$ times in any order, we have $2^{d-1}$ plays each of which with $10^d$ different instantiations.
The total number of symbolic plays is $1+2^1+2^2+\ldots+2^{d-1} = 2^d-1$.

In general, for a play $p \in \sbr{\Gamma \vdash_{d} M:T}$ where $M$ contains $m \geq 0$ calls to
an undefined function with $n$ arguments, we have:
\begin{equation} \label{eq:undef-fun}
\#(ID_p) = (1+n+n^2+\ldots+n^{d-1})^m \cdot {\textstyle \prod_{Z \in p} } \! |\!dom(Z)\!|, \quad  \#(pc_p) \leq {\textstyle \prod_{Z \in p} } \! |\!dom(Z)\!|
\end{equation}
Note that if the undefined function has 1 argument, then the total number of symbolic plays is $1+1^2+\ldots+1^{d-1}=d$.
When the play condition is $\true$, then $\#(\true_p) = \prod_{Z \in p} |\!dom(Z)\!|$.

\begin{example}\label{exp:fun}
Consider the term:
\[
\begin{array}{l}
f : \mathsf{com}^{f,1} \to \mathsf{expint}_{10}^{f}, \mathsf{abort}:\mathsf{com}^{abort} \vdash_5 \ \mathsf{new_{int}} \, x:=0 \ \mathsf{in} \\
\qquad \qquad \qquad \qquad \qquad \qquad \qquad \qquad \quad \mathsf{if} \, (f(x:=!x+1)+!x>3) \ \mathsf{then} \ \mathsf{skip} \ \mathsf{else} \ \mathsf{abort} : \mathsf{com}
\end{array}
\]
where we bound the size of context on definitions of ``$f$'' which can call its argument at most 4 times.
Note that ``$f$'' has 1 argument and is called once in the above term.
The symbolic model of the above term is:
\[
\begin{array}{l}
\ibr{?X\!=\!0,\sr} \cdot \sq^{f} \cdot \sum_{k=0}^{4} \big(\ibr{?X\!=\!X\!+\!1, \sr^{f,1}} \cdot \sd^{f,1} \big)^k \cdot ?Z^f \cdot \\ \qquad \qquad \qquad
\big( \ibr{Z\!+\!X\!>\!3, \sd} + \ibr{Z\!+\!X\!\leq\!3, \sr^{\sa}} \cdot \sd^{\sa} \cdot \sd \big)
\end{array}
\]
For the contexts corresponding to ``$f$'' which does not call its argument at all ($X=0$), the unsafe behaviour is exercised
when the value returned from $f$ is $Z \in [0,4)$, i.e.\ the failure probability is $(1/5)\cdot(4/10)$.
When the function ``$f$'' calls its argument once, the variable $x$ is incremented once ($X=1$) and so the failure probability is $(1/5)\cdot(3/10)$.
For the contexts when ``$f$'' calls its argument twice ($X=2$), $\mathsf{abort}$ is run with the likelihood
$(1/5)\cdot(2/10)$; when ``$f$'' calls its argument three times ($X=3$) the failure probability is $(1/5)\cdot(1/10)$; whereas
when ``$f$'' calls its argument four times ($X=4$), the failure probability is $0\%$ ($Z+X\!\leq\!3$
is unsatisfiable).
Therefore, 
for $d=5$, the failure
probability is $(4/50)+(3/50)+(2/50)+(1/50)+(0/50)=10/50 \, (20\%)$; whereas the success probability is $40/50 \, (80\%)$.

When $d=6$, the failure probability is $10/60 \, (16.7\%)$, and the success is $83.3\%$.
For $d=10$, the failure is $10/100 \, (10\%)$, and the success is $90\%$.
\end{example} 

\section{Implementation} \label{sec:impl}
We have extended the \textsc{Symbolic GameChecker} tool \cite{D14} to implement our approach for
performing probabilistic analysis of open terms.
The basic tool \cite{D14} converts any IA$_2$ term into a symbolic automaton representing its game semantics, and
then explores the automaton for unsafe traces (plays).
It calls an external SMT solver, Yices \cite{DBLP:conf/cav/Dutertre14}, 
to determine satisfiability of play conditions.
The extended tool performs a bounded probabilistic analysis on the obtained symbolic automaton
in order to determine the success and failure probabilities of the input term.
Instead of an SMT solver, the extended tool calls a model counter, \textsc{LattE} \cite{LattE}, to
determine the number of solutions to play conditions.
We now illustrate our tool with an example.
The tool, further examples and reports on how they execute are available from:
\verb|https://aleksdimovski.github.io/symbolicgc.html| (version for probabilistic analysis).

Consider the following version of the linear search algorithm:
\begin{tabbing}
 $x[k] \, : \, \mathsf{varint}_{n}^{x[-]}, \ y \, : \, \mathsf{expint}_{n}^{y}, \ \mathsf{abort} \, : \, \mathsf{com}^{abort} \ \vdash$ \\
 $\qquad \mathsf{new}_{int} \, i \aasg 0 \, \mathsf{in} $ \\
 $\qquad \mathsf{new}_{int} \, p \aasg y \, \mathsf{in} $ \\
 $\qquad \mathsf{while} \, (i<k) \, \mathsf{do} \, \{$ \\
 $\qquad \qquad \mathsf{if} \, (x[i] = p) \, \mathsf{then} \ \mathsf{abort};$ \\
 $\qquad \qquad i := i + 1; $ \\
 $\qquad  \} \, : \mathsf{com}$
\end{tabbing}
The meta variable $k>0$ represents the size of array $x$, and $n>0$ represents
the domain size of input expressions $y$ and $x[0], \ldots, x[k-1]$.
Both $k$ and $n$ will be replaced by several different values.
In the above term, first the input expression $y$ is copied into the local variable $p$.
Then the non-local array $x$ is searched for an occurrence of the value stored in $p$.
If the search succeeds, $\mathsf{abort}$ is executed.
We assume that $y$ and all elements of the array $x$ can take one uniform value
from the range $[0,n)$, i.e.\ their type is $\mathsf{expint}_{10}$.

\begin{figure*}[t]
\begin{center}
\begin{tikzpicture}[->,>=stealth',shorten >=1pt,auto,node distance=1.5cm,
                    semithick]
  \tikzstyle{every state}=[minimum size=.3pt,initial text={{\scriptsize start}}]

  \node[initial below,state] (A)                    {\tiny{$0$}};
  \node[state]         (B) [right of=A] {\tiny{$1$}};
  \node[state]         (F) [right of=B]  {\tiny{$2$}};
  \node[state]         (N) [right of=F]  {\tiny{$3$}};
  \node[state]         (C) [below right of=N] {\tiny{$4$}};
  \node[state,,accepting]         (D) [below left of=N, node distance=19mm]  {\tiny{$5$}};
  \node[state]         (L) [below of=C, node distance=16mm]  {\tiny{$6$}};
  \node[state]         (O) [below of=L]  {\tiny{$7$}};
  \node[state]         (P) [left of=O, node distance=19mm]  {\tiny{$8$}};

\path[font=\scriptsize,dashed]
        (A) edge              node {\tiny{$\ibr{?I\!\!=\!\!0,\sr}$}} (B)
        (F) edge              node {\tiny{$?Y^y$}} (N)
        (C) edge              node[above,sloped] {\tiny{$?Z^{x[I]}$}} (L)
        (O) edge              node {\tiny{$\sd^{abort}$}} (P);
  \path[font=\scriptsize]
        (B) edge              node {\tiny{$\sq^y$}}  (F)
        (N) edge    [bend left]          node[right] {\tiny{$\ibr{?P=Y\!\land\!I < k, \srd^{x[I]}}$}} (C)
        (N) edge    [bend right]          node[left] {\tiny{$\ibr{?P=Y\!\land\!I \geq k, \sd}$}} (D)
        (L) edge    [bend left]          node[above,left] {\tiny{$\ibr{P \neq Z\!\land\!?I\!=\!I\!+\!1\!\land\!I\!\geq\!k, \sd}$}} (D)
        (L) edge    [bend right=60]          node[above,right] {\tiny{$\ibr{P \neq Z\!\land\!?I\!=\!I\!+\!1\!\land\!I\!<\!k, \srd^{x[I]}}$}} (C)
        (L) edge              node {\tiny{$\ibr{P=Z, \sr^{abort}}$}} (O)
        (P) edge              node {\tiny{$\ibr{?I\!=\!I\!+\!1\!\land\!I \geq k, \sd}$}} (D)
        (P) edge     [bend left=30]         node[below,sloped] {\tiny{$\ibr{?I\!=\!I\!+\!1\!\land\!I\!<\!k, \srd^{x[I]}}$}} (C)
       ;
\end{tikzpicture}
\end{center}
\vspace{-2mm}
\caption{The model for the linear search term. } \label{fig:linear}
\end{figure*}
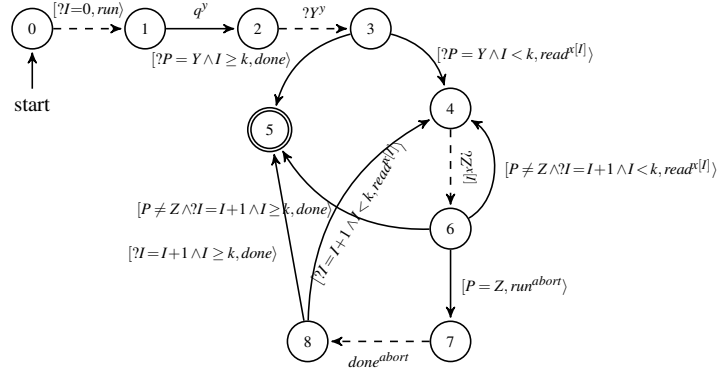

The symbolic model for this term is given in Fig.~\ref{fig:linear}.
The array $x[k]$ is given a symbolic representation \cite{D14}, where the array size $k$ and the index of the array elements represent symbols.
We use symbols $I$ and $P$ to track the current values of local
variables $i$ and $p$, respectively. The symbol $I$ is used to represent the index of an array element
that needs to be de-referenced or assigned to.
If the value $Y$ read from the environment \textbf{O} for the expression $y$ is equal to the value
$Z$ read from the environment \textbf{O} for some array element $X[I]$, where $0 \leq I < k$, i.e.\ the
constraint $(P=Y) \!\land\! (P=Z)$ holds, then an unsafe behaviour is exercised.

We now present the probabilistic analysis of the above term for $n=10$ and for various concrete values of $k$.
The exploration bound is $d=k-1$.
For $k=1$, $\mathsf{abort}$ is run only when the values $Y \in [0,10)$ and
$Z \in [0,10)$ read from the environment for $y$ and $x[0]$, respectively, are equal.
Hence, the failure probability is $10\%$, and the success probability is $90\%$.
For $k=2$, we obtain only one feasible safe trace when $Y \neq Z_1$ and $Y \neq Z_2$ for the values $Y, Z_1, Z_2 \in [0,10)$
read from the environment for $y$, $x[0]$ and $x[1]$, respectively.
Therefore, the success probability is $(9 \cdot 9) / (10 \cdot 10)=81\%$, and the failure probability is $19\%$.
For $k=3$, we obtain that the success probability is
$72.9\%$ and the failure probability is $27.1\%$.
For $k=5$, we have
$Pr^s=59.05\%$ and $Pr^f=40.95\%$.
For $k=10$ and $d=9$, the success probability is $40.7\%$ and the failure probability is $59.3\%$.
We notice that as the array size $k$ grows, the likelihood of running $\mathsf{abort}$ grows as well
since there are more array elements in this case and the probability that some of them is equal to
the input expression $y$ is bigger.

We now report experimental results for performing the probabilistic analysis of the linear search
term for different values of $k$ (size of array) and $n$ (domain size of undefined expressions).
We ran our tool on a 64-bit Intel$^\circledR$Core$^{TM}$ i5 CPU and 8 GB memory.
The performance numbers reported constitute the average runtime of five independent executions.

The symbolic model has 9 states and the total time needed to generate the model is $0.24$ sec.
Note that the model size and the time needed to generate it are the same for all values of $k$ and $n$.
The results from running probabilistic analysis for this term are shown in Table~\ref{table:exp}.
For different values of $k$ 
we list:
the number of generated unsafe and safe traces, 
the total number of visited (re-explored) states during the analysis, and
 the execution time in seconds
needed to perform the analysis (search) when $n=10$, $n=256$ (1 byte) and $n=65,536$ (2 bytes).
We perform three sets of experiments, the first when the domain size of undefined
 expressions is $n=10$, the second when the domain size is $n=256$, and the third when $n=65,536$.
 We only show the different analysis times corresponding to various values of $n$, since the first three parameters
 are the same in all cases.
 We observe that we obtain similar time performance results for $n=10$, $n=256$, and $n=65,536$, mostly due to the fact that \textsc{LattE}
 is largely insensitive to those values in terms of time.
 On the other hand, the analysis time increases for bigger values of $k$.
 In those cases, we have more traces to analyze and more complex constraints, which lead to more calls to \textsc{LattE}.

\begin{table}[t]
\begin{center}
{\renewcommand{\arraystretch}{1}
\begin{tabular}{||c||c|c|c|c|c|c||}  \hhline{|t:=t:======:t|}
  \multirow{2}{*}{~~$k$~~}  &  \multicolumn{2}{c|}{$\#$ traces}  & \multirow{2}{*}{~$\#$ states~} &  \multicolumn{3}{c||}{Analysis Time}  \\
\hhline{~--~---}
     &  unsafe  &  ~~safe~~  &   &  ~$n=10$~ & ~$n=256$~ & ~$n=65,536$ \\
 \hhline{|:=::======:|}
 1 & 1 & 2  & 35  & 0.83 & 0.87   &   0.99    \\ \hhline{||-||------||}
 3  & 11 &  4  &  162  & 6.43 & 6.55   & 6.83 \\ \hhline{||-||------||}
 5 &  57 & 6   &  756  & 52.50  &  55.52  & 56.55 \\ \hhline{||-||------||}
 6  & 120 &  7 &  1677  & 166.58 &   169.02   &  175.32       \\ \hhline{||-||------||}
 7 & 247 &  8  &  3758  & 607.40  & 611.35  &  619.14    \\ \hhline{|b:=:b:======:b|}
\end{tabular}}
\caption{Performance of the probabilistic analysis of the linear search term
for different values of the array size $k$ and the domain size $n$. Time is in seconds (s).
}\label{table:exp}
\end{center}
\end{table} 


\section{Related work}\label{sec:related}

Traditional formal approaches for probabilistic analysis based on probabilistic model checking \cite{DBLP:conf/tacas/KwiatkowskaNP02}
require a high-level design of the software.
However, such models are difficult to maintain and may abstract important details that impact the chance
of property satisfaction in the system.
The ultimate goal is to perform probabilistic analysis directly on implementations, not on high-level models.
Recent approaches \cite{DBLP:conf/issta/GeldenhuysDV12,DBLP:conf/icse/FilieriPV13,DBLP:conf/pldi/BorgesFdPV14} have proposed to use symbolic execution to support probabilistic
analysis on the source code.
In this work, we define probabilistic analysis in the settings of game semantics.
This brings several distinctive features to our approach, such as: very precise models, compositional
modelling, models of open second order imperative programs with free identifiers which take into account all possible contexts
in which the programs can be considered.
In contrast, the works in \cite{DBLP:conf/issta/GeldenhuysDV12,DBLP:conf/icse/FilieriPV13,DBLP:conf/pldi/BorgesFdPV14}
consider imperative programs that only have some undefined global variables.

Game semantics for full Idealized Algol has been defined before \cite{DBLP:journals/entcs/AbramskyM96}.
The applications to software model checking were first proposed in \cite{GM}, where
game semantics models for second-order IA with finite data types were represented as
 finite automata. 
 By using symbols instead of concrete data for inputs,
it was shown \cite{DBLP:journals/corr/abs-1210-2454,D14} how to generate finite symbolic automata for second-order IA with infinite data types.
We have shown how to extend symbolic automata in order to represent program families implemented
using \texttt{\#ifdef} annotations \cite{DBLP:conf/spin/Dimovski16}.
Specifically designed model checking algorithms are then employed to verify safety of all variants of the family at once,
 and report those variants that are unsafe (resp., safe).
 Algorithmic game semantics also provides a method \cite{DBLP:journals/corr/Dimovski13,DBLP:conf/stm/Dimovski14} for ensuring secure information flow of open programs, i.e.\ for verifying
security properties such as timing leaks, non-interference, and termination leaks.
A fully abstract game semantics models for Probabilistic Idealized Algol (PA) has been formally defined in \cite{DBLP:journals/tocl/DanosH02}.
PA extends IA by allowing (fair) coin-tossing as a valid expression.
Algorithmic probabilistic game semantics for PA
have been studied as well \cite{DBLP:conf/concur/MurawskiO05,DBLP:journals/fmsd/KieferMOWW13}.
In particular, probabilistic equivalence and refinement have been explored
in the context of game semantics.
Intuitively, two probabilistic programs are equivalent if for each input they give rise to identical probabilistic distributions on the set of possible outputs.
An automated equivalence checker for PA is also
developed which takes a program as input and returns a probabilistic automaton capturing the game semantics
of the program.
An interesting direction for future work would be to calculate path probabilities and program
reliability for PA terms.

\section{Conclusion}\label{sec:conclusion}

In this work, we show how game semantics and model counting can be used to give a specific quantitative
analysis of open programs -- calculation of program path probabilities.
We also apply the obtained analysis results for predicting program reliability.

Our analysis used the IA$_2$ language in order to stay focused, but a similar approach can be extended
to any language for which an algorithmic game semantics exists.
The model considered here contains only convergent behaviours of terms, and so it is suitable for
verifying safety properties. If we want to take into account liveness properties as well, then the model
should be enriched to contain all possible divergent behaviors of terms \cite{DBLP:conf/lics/HarmerM99,DBLP:conf/ifm/Dimovski10}.
For such models, we can calculate probabilities for convergence and divergence of terms similarly as
we did here for success and failure.




\bibliographystyle{eptcs}
\bibliography{main-bib}

\end{document}